\documentclass[a4paper,11pt]{article}
\usepackage{pos}
\usepackage{bm}
\usepackage{slashed}

\newcommand{\tp}{\text{p}}

\title{Intrinsic quantum coherence in particle oscillations}
\author*{Anca Tureanu}

\affiliation{Department of Physics, University of Helsinki, \\
P.O.Box 64, FI-00014 Helsinki, Finland}

\emailAdd{anca.tureanu@helsinki.fi}

\abstract{The quantum field theoretical description of coherence in the oscillations of particles, especially neutrinos, is a standing problem in particle physics. In this talk, several inconsistencies of 
the standard approach to particle oscillations will be explained, and 
how they are resolved in a process-independent manner, by a novel approach inspired by the 
Bardeen--Cooper--Schrieffer theory of superconductivity and the Nambu--Jona-Lasinio model. The formalism leads to corrections to the neutrino oscillation probability originally written by Pontecorvo and Gribov, however the standard probability is validated in the ultrarelativistic neutrino limit. The massive neutrino states are interpreted as quasiparticles on a vacuum condensate of "Cooper pairs" of massless flavour neutrinos. The newly defined oscillating particle states are for neutrino oscillations what the Klauder--Sudarshan--Glauber coherent states are for quantum optics.
}

\FullConference{%
  40th International Conference on High Energy physics - ICHEP2020\\
  July 28 - August 6, 2020\\
  Prague, Czech Republic (virtual meeting)
}

\begin{document}
\maketitle

\section{Introduction}

The phenomenon of particle oscillations, epitomized by the neutrino oscillations, is the most spectacular realization of quantum coherence in Nature. It is believed that the phenomenon has two possible descriptions, namely a quantum mechanical and a quantum field theoretical one. Below we shall argue that only the latter description is conceptually plausible: in spite of the formal analogies, the particle oscillations are not a quantum mechanical process! However, over 60 years from the theoretical prediction and over 20 years from the experimental observation of the neutrino oscillations, we have to face the lack of a consistent and widely-accepted quantum field theoretical formulation of the mixing and oscillation phenomenon. Even the most recent monographs and textbooks on quantum field theory do not include a chapter on particle oscillations.  

To date, the most compeling evidence for the never yet measured neutrino masses are the neutrino oscillations. In the paradigm of Pontecorvo \cite{Pontecorvo2,Pontecorvo3}, the flavour neutrino states produced and detected in weak interactions are {\it coherent superpositions} of massive neutrino states with different masses and therefore able to oscillate from one flavour to another. These ideas emerged when no experimental hint of neutrino oscillations existed, based on the theoretical work of Gell-Mann and Pais regarding the strangeness violating $K_0-\bar K_0$ oscillations (for a review, see \cite{Bilenky_hist}). Later on, the possible baryon number violating neutron-antineutron oscillations have been also hypothesized. While the mesons and baryons, as composite particles, are subject to several interactions and the oscillation is perhaps intuitively easier to picture, the elementary neutrinos with their unique weak interactions are the most interesting case.

What precludes oscillating neutrinos from fitting into the conventional quantum field theory? The reasons are two-fold: i) their coherence is intrinsic, i.e. irrespective of any trigger and not controllable in the experiments; ii) only three massive neutrino states can form a coherent superposition. The theory of coherence pioneered by Klauder \cite{Klauder}, Sudarshan \cite{Sudarshan} and Glauber  \cite{Glauber}, based on the notion of coherent states and customarily used in quantum optics, does not cover so far the {\it intrinsic quantum coherence} of neutrinos.

In the following, we shall briefly discuss the basic inconsistencies of a quantum mechanical treatment and review a recent proposal of the author for tackling in a quantum field theoretical framework the construction of flavour states with inbuilt quantum coherence \cite{AT, AT-neutron}.

\section{Standard description of neutrino oscillations}
The basis of the standard description is the addition of flavour violation terms (induced by Yukawa terms) to the Lagrangian of the SM \cite{Bilenky_hist}. The quadratic part of the Lagrangian is diagonalized
 in terms of massive neutrino fields $\Psi_{1},\Psi_{2}$ of masses $m_1,m_2$, in a two-neutrino mixing model:
\begin{eqnarray}\label{rotation}
\left(\begin{array}{c}
            \Psi_{\nu_e}(x)\\
            \Psi_{\nu_{\mu}}(x)
            \end{array}\right)= \left(\begin{array}{c c}
            \cos\theta &\sin\theta\\
           -\sin\theta&\cos\theta
            \end{array}\right)\left(\begin{array}{c}
            \Psi_{1}(x)\\
            \Psi_{2}(x)
            \end{array}\right),\ \ \ \ \ \  \tan^2\theta=\frac{2m_{e\mu}}{m_{\mu\mu}-m_{ee}}.
\end{eqnarray}
The fields $\Psi_{\nu_e}$ and $\Psi_{\nu_\mu}$ are called {\it flavour neutrino fields } (though flavour is violated). Let us remark that they are {\it interacting fields}, for which one cannot define creation and annihilation operators, therefore no Fock states.

%The canonical quantization of the diagonalized Lagrangian is trivial
%
%$$\Psi_i(x)=\int\frac{d^3 p}{(2\pi)^{3/2}\sqrt{E_{ip}}}\sum_\lambda\left(A_{i\lambda}({\bf p})U_\lambda({\bf p})e^{-ipx}+B^\dagger_{i\lambda}({\bf p})V_\lambda({\bf p})e^{ipx}\right),\ \ i=1,2$$
%and we obtain the massive neutrino  and antineutrino states: $|\nu_{i\lambda}({\bf p})\rangle=A^\dagger_{i\lambda}({\bf p})|\Phi_0\rangle,\ \ \ |\bar\nu_{i\lambda}({\bf p})\rangle=B^\dagger_{i\lambda}({\bf p})|\Phi_0\rangle$, where $|\Phi_0\rangle$ is the physical vacuum.

According to Pontecorvo's conjecture, one can {\it define} flavour neutrino states $|\nu_e\rangle,|\nu_\mu\rangle$ as coherent superpositions of the massive neutrino states $|\nu_1\rangle,|\nu_2\rangle$ with different masses $m_1,m_2$, by replicating the mixing formula for the fields \eqref{rotation}:
\begin{eqnarray}
\left(\begin{array}{c}
            {|{\nu_e}\rangle}\\
            {|{\nu_{\mu}}\rangle}
            \end{array}\right)= \left(\begin{array}{c c}
            \cos\theta &\sin\theta\\
           -\sin\theta&\cos\theta
            \end{array}\right)\left(\begin{array}{c}
            {|\nu_1\rangle}\\
            {|\nu_2\rangle}
            \end{array}\right).
\end{eqnarray}
Then the oscillations can take place with the probability:
\begin{eqnarray}\label{osc_p}{\cal P}_{\nu_e\to\nu_\mu}=|\langle\nu_{\mu}({\bf p})|e^{-iHt}|\nu_{e}({\bf p})\rangle|^2
=\sin^2(2\theta)\sin^2\left(\frac{{\Delta m^2}}{4E}L\right),\ \ \  
{\Delta m^2}={m_2^2-m_1^2},\ \ \frac{m_i}{E}\ll 1.
\end{eqnarray} 

The requirements for neutrino oscillations are: 1) a flavour-violating Lagrangian; 2) massive neutrinos; 3) flavour neutrino states as coherent superpositions of massive neutrino states with different masses (belonging to different Fock spaces).

 Let us recall, however, that in QFT particles with different masses are always {\it incoherently} produced and absorbed!

There have been many attempts to incorporate the oscillation phenomenon into quantum field theory (for a review, see \cite{Akh-Kopp}), but the conclusion is, as expected, that {\it coherent flavour neutrino states cannot be derived in the conventional framework of QFT}.

\section{Oscillation of states and coherence in Quantum Mechanics}
The prototype for the above description of neutrino oscillations is the quantum mechanical two-level system. This is a system with two stationary states, that get mixed by the sudden switching on of an interaction.

Assume that the system is initially described by Hamiltonian $H_0$ with the orthonormal basis states $|0\rangle$ and $|1\rangle$. The system prepared in the stationary state $|0\rangle$ and it evolves with $H_0$. At 
$t=t_0$, the interaction is turned on {\it suddenly (diabatically)} and the new Hamiltonian is
$H=H_0+ H_{int}$, with
a new basis of stationary states $|\phi_1\rangle$ and $|\phi_2\rangle$, i.e. $
H|\phi_i\rangle=E_i|\phi_i\rangle, \ \ i=1,2.
$
The diabaticity of the process makes so that the system does not transition slowly into a stationary state of $H$, but it remains in the state $|0\rangle$, while evolving with the Hamiltonian $H$. The initial state $|0\rangle$ is a {\it coherent superposition} of the states $|\phi_1\rangle$ and $|\phi_2\rangle$,
\begin{equation}|0\rangle=c_1|\phi_1\rangle+c_2|\phi_2\rangle, \ \ \ |c_1|^2+|c_2|^2=1.\end{equation}

At the time $t=t_0+\Delta t$ we remove suddenly the interaction and determine the state of the system, which can be either $|0\rangle$ or $|1\rangle$. The probability that the system has transitioned into the state $|1\rangle$ is:
\begin{equation}{\cal P}_{|0\rangle\to|1\rangle}=|\langle 1|e^{-iH\Delta t}|0\rangle|^2\sim \sin^2\left(\frac{\Delta E}{2}\Delta t\right).\end{equation}

The question is whether the above simple picture can be applied to particle oscillations. Let us note the following facts regarding the two-level system:
i) due of the Stone--von Neumann theorem, all the representations of the canonical algebra for a given quantum mechanical system are equivalent, implying a {\it unitary change of basis}:
\begin{eqnarray}\label{rotation_modes}
\left(\begin{array}{c}
           |0\rangle\\
           |1\rangle
            \end{array}\right)= \left(\begin{array}{c c}
            c_1 &c_2\\
           c_3&c_4
            \end{array}\right)\left(\begin{array}{c}
            |\phi_1\rangle\\
            |\phi_2\rangle
            \end{array}\right);
\end{eqnarray}
ii) the states of the two bases are well-defined as stationary states of either $H_0$ or $H$; iii) the coherent superposition of states (leading to interference and finally to oscillation) is achieved by {\it turning on/off suddenly the interaction.}

In the case of neutrinos, on the other hand: i) the flavour violating part of the Lagrangian (mixing the flavour fields) cannot be turned on and off at will; ii) the coherence of flavour neutrino states is not triggered by external factors, it is {\bf intrinsic}; iii) the quantum mechanical principle of superposition of states fails: {\it the two massive neutrino states which are superposed are not states of the same system, but states of two distinct systems}!

Consequently, in spite of the formal similarities, the quantum mechanical interpretation of neutrino oscillations as two-level system oscillations is conceptually untenable.

\section{Coherent states in quantum optics}
A related concept of coherence appears in the formulation of {\it coherent states} \cite{Klauder, Sudarshan,Glauber}, used mainly in quantum optics. They are eigenstates of the annihilation operator of the harmonic oscillator,
$\hat a|\alpha\rangle= \alpha|\alpha\rangle,\ \ \hat a|0\rangle=0,$
where
$\alpha=|\alpha|e^{i\theta}$ is a complex number. The states satisfying the above are
\begin{equation}\label{coherent}|\alpha\rangle= e^{\alpha\hat a^\dagger-\alpha^*\hat a}|0\rangle=e^{-\frac{|\alpha|^2}{2}}\sum_{n=0}^\infty\frac{\alpha^n}{\sqrt{n!}}|n\rangle,\end{equation}
i.e. the coherent states are superpositions of {\it an infinite number of particle states} (or Fock states), all belonging to {\it the same Fock space}. In QFT, a similar notion of coherent state appears as vacuum condensate.

In this framework, the question is: How to define coherent oscillating states in quantum field theory, as superposition of a {\it finite number of particle states} belonging to {\it different Fock spaces}?

\section{Intrinsically coherent oscillating particle states}
 Returning to first principles, let us recall that in QFT, particle states are defined by the action of an operator on the physical vacuum of the theory. Though they are not Fock space states, the coherent states \eqref{coherent} are defined by the same principle. We have proposed to adopt the same guiding line for defining intrinsically coherent oscillating particle states \cite{AT,AT-neutron}. In the case of neutrinos, this idea leads to associating the oscillating neutrino states to the massless flavour neutrino fields of the Standard Model \cite{AT}.

To this end, one connects massless to massive neutrino fields by a procedure reminiscent of the Nambu--Jona-Lasinio model for dynamical generation of nucleon masses \cite{NJL}, inspired by the Bardeen--Cooper--Schrieffer (BCS) theory of superconductivity in Bogoliubov's formulation \cite{Bogoliubov}. The technique is described in detail for Dirac neutrino oscillations in \cite{AT} and for neutron-antineutron oscillations in \cite{AT-neutron}, therefore we shall mention here only the essential features and results. The idea is to regard all the mass terms, including the flavour mixing terms in the Lagrangian, as interaction terms between massless neutrino fields. This picture is warranted by the construction of the SM, in which all the mass terms for fermions arise dynamically by the attractive Yukawa interactions with the Higgs field.

Specifically, one expands the quadratic Hamiltonian of the flavour violating theory, in Schr\"odinger picture, in terms of the canonical creation and annihilation operators of massless flavour fields acting on the states of a Fock space whose vacuum is $|0\rangle$. Naturally, the Hamiltonian is nondiagonal
 in terms of massless (bare) particles' operators $a_{e\lambda},  a_{\mu\lambda}, b_{e\lambda}, b_{\mu\lambda}$:

The diagonal form of the Hamiltonian
%\begin{eqnarray}\label{H_osc_diag}
%{\color{magenta}H=\int d^3{ p}\sum_{\lambda,i=1,2} E_{i\tp} \Big[A^\dagger_{i\lambda}({\bf p}) A_{i\lambda}({\bf p})+ B^\dagger_{i\lambda}({\bf p}) B_{i\lambda}({\bf p})\Big],}\ \ \ \ E_{i\tp}=\sqrt{\tp^2+m_i^2}
%\end{eqnarray}
is achieved by two canonical transformations and
the eigenstates of the diagonal Hamiltonian are the physical particle states (Bogoliubov quasiparticles), specifically the massive neutrino states. They are elements of a new Fock space, with the physical vacuum $|\Phi\rangle$.

The two canonical transformations are:

- unitary transformation (rotation) between the operators of the massless fields:
\begin{eqnarray}\label{rotation_modes}
\left(\begin{array}{c}
            {a_{e\lambda}({\bf p})}\\
             {a_{\mu\lambda}({\bf p})}
            \end{array}\right)= \left(\begin{array}{c c}
            \cos\theta &\sin\theta\\
           -\sin\theta&\cos\theta
            \end{array}\right)\left(\begin{array}{c}
            a_{1\lambda}({\bf p})\\
            a_{2\lambda}({\bf p})
            \end{array}\right);
\end{eqnarray}

- Bogoliubov transformations between the "massless" and "massive" operators ($A_{i\lambda},B_{i\lambda}$):
\begin{eqnarray}\label{BT}
{A_{i\lambda}({\bf p})}&=&\alpha_{i\tp}a_{i\lambda}({\bf p})+\beta_{i\tp}b^\dagger_{i\lambda}(-{\bf p}),\ \  \ \  i=1,2,\cr
{B_{i\lambda}({\bf p})}&=&\alpha_{i\tp}b_{i\lambda}({\bf p})-\beta_{i\tp}a^\dagger_{i\lambda}(-{\bf p}),\ \ 
\end{eqnarray}
where
$$\alpha_{i\tp}=\sqrt{\frac{1}{2}\left(1+\frac{\tp}{E_{i\tp}}\right)},\  \ \ \ 
\beta_{i\tp}=\text{sgn}\,\lambda\,\sqrt{\frac{1}{2}\left(1-\frac{\tp}{E_{i\tp}}\right)}.$$

The physical vacuum is a condensate of "Cooper-like pairs" of massless neutrino-antineutrino, therefore a coherent state in the spirit of the previous section:
\begin{eqnarray}\label{normalized vacuum}
|\Phi_0\rangle =\ \Pi_{i,{\bf p},\lambda}\  \left(\alpha_{i\tp} -\beta_{i\tp}\,a_{i\lambda}^\dagger({\bf p})b_{i\lambda}^\dagger(-{\bf p})\right)|0\rangle,
\end{eqnarray}
such that
$
{\langle 0|\Phi_0\rangle  }= \Pi_{i,{\bf p},\lambda}\ \alpha_{i\tp}=\Pi_{i,{\bf p},\lambda}\ \left(1+\frac{\tp}{E_{i\tp}}\right)^{1/2}\to \exp\left[-(m_1^2+m_2^2)\int d{ \tp}\right]=0
$
in the infinite volume and momentum limit.
The Fock spaces built on the vacua $|0\rangle$ and $|\Phi_0\rangle$ do not contain any common states.

We can now define inherently coherent oscillating neutrino states by the prescription:
\begin{eqnarray}\label{nu_osc}
|\nu_e({\bf p},\lambda)\rangle&\equiv&{a_{e\lambda}^\dagger({\bf p})|\Phi_0\rangle}=\cos\theta{\sqrt{1/2+{\tp}/{2E_{1\tp}}}}|\nu_{1}({\bf p})\rangle+\sin\theta{\sqrt{{1}/{2}+{\tp}/{2E_{2\tp}}}}|\nu_{2}({\bf p})\rangle,\\
|\nu_\mu({\bf p},\lambda)\rangle&\equiv&{a_{\mu\lambda}^\dagger({\bf p})|\Phi_0\rangle}=-\sin\theta{\sqrt{{1}/{2}+{\tp}/{2E_{1\tp}}}}|\nu_{1}({\bf p})\rangle+\cos\theta{\sqrt{{1}/{2}+{\tp}/{2E_{2\tp}}}}|\nu_{2}({\bf p})\rangle.\nonumber
\end{eqnarray}
The corresponding oscillation amplitude is never zero:
$$
{\cal A}_{\nu_e\to\nu_\mu}(t)=\frac{1}{2}\sin 2\theta e^{-i\tp t}\Big[-\left(1-\frac{1}{4}\frac{m_1^2}{\tp^2}\right)^2e^{-i\frac{m_1^2}{2\tp}t}
+\left(1-\frac{1}{4}\frac{m_2^2}{\tp^2}\right)^2e^{-i\frac{m_2^2}{2\tp}t}\Big],
$$
meaning that there is always a portion (though negligible for relativistic particles) of muon neutrino in the electron neutrino and vice-versa. The lack of orthogonality is specific to coherent states. In the ultrarelativistic limit, one recovers Pontecorvo's oscillation probability \eqref{osc_p}.

\section{Conclusions and outlook}
The coherence of flavour states is the key element for oscillations, which cannot be implemented by usual QFT prescriptions.
The proposed construction of {\it intrinsically coherent neutrino states} \eqref{nu_osc} is based on establishing a one-to-one correspondence with the Standard Model massless neutrino states \cite{AT,AT-neutron}. The procedure of defining oscillating particle states can be implemented for any type of oscillating systems ($K_0-\bar K_0$, $n-\bar n$, Majorana neutrinos, seesaw mechanism).
Though the Pontecorvo oscillation probability is obtained in the ultrarelativistic limit, there are quantitatively significant differences for nonrelativistic neutrinos (see KATRIN and PTOLEMY experiments) and possibly for Mikheyev--Smirnov--Wolfenstein effect in matter (especially for neutrinos in extreme conditions).

This is the first step towards elucidating 
the mechanism of interaction (production and absorbtion) of oscillating particle states.

%
%% Commands for Bibliography:
\newcommand{\atitle}[1]{\emph{#1},}
\newcommand{\jref}[2]{\href{https://doi.org/#2}{#1}}
\newcommand{\arXiv}[2]{\href{http://arxiv.org/abs/#1}
{\texttt{arXiv:#1 [#2]}}}
\newcommand{\arXivOld}[1]{\href{http://arxiv.org/abs/#1}{\texttt{#1}}}

\end{document}